\documentclass{ws-procs975x65}

\begin{document}

\title{LOOKING BEYOND THE HORIZON}

\author{EUGENY BABICHEV}

\address{INFN - Laboratori Nazionali del Gran Sasso, S.S.
 17bis, 67010 Assergi (L'Aquila), Italy\newline 
and Institute for Nuclear Research of the Russian Academy of Sciences, 60th
October Anniversary Prospect 7a, 117312 Moscow, Russia\\
\email{babichev@lngs.infn.it}}

\author{VIATCHESLAV MUKHANOV}

\address{Arnold-Sommerfeld-Center for Theoretical Physics, Department f\"ur
Physik, Ludwig-Maximilians-Universit\"at M\"unchen, Theresienstr. 37,
D-80333, Munich, Germany\newline 
and Perimeter Institute for Theoretical Physics, 
31 Caroline Street North Waterloo, Ontario, N2L 2Y5, Canada\\
\email{mukhanov@theorie.physik.uni-muenchen.de}}

\author{ALEXANDER VIKMAN}

\address{Arnold-Sommerfeld-Center for Theoretical Physics, Department f\"ur
Physik, Ludwig-Maximilians-Universit\"at M\"unchen, Theresienstr. 37,
D-80333, Munich, Germany\\
\email{vikman@theorie.physik.uni-muenchen.de}}

\bodymatter


In our previous work \cite{we} we have shown that it is principally possible to send information from the interior of a black hole (BH) without modifying Hilbert-Einstein action \footnote{In bimetric theories this possibility can be easily achieved \cite{BiM}.}. This may happen provided a special type of scalar field theory with noncanonical kinetic term is realized in nature. In our approach we have considered the accretion of a particular noncanonical field $\phi$ onto a black hole \footnote{The accretion of other noncanonnical scalar fields were studied in e.g. \cite{Mukohu,BDE}.}. The scalar field flow forms an acoustic black hole similar to the well known "dumb hole" \cite{Unruh}. This background dynamically breaks Lorentz invariance and serves as a "new ether". Due to the nonlinearity of the equation of motion, the "new ether" allows a superluminal propagation of small perturbations $\delta \phi$ of the field $\phi$ \footnote{or of other kinetically coupled to $\phi$ fields as in \cite{DubovSib}. In this case it is also possible to get information from the interior of BH.}. The field theories with nonstandard kinetic terms have been subjects of investigation since a long time ago \cite{BI}. In cosmology they were first introduced in the context of k-inflation and k-essence models \cite
{k-inflation}. The letter seems to require a superluminal sound speed during a period of cosmic evolution \cite{Durrer}. The models with superluminal sound speed may have other interesting applications in cosmology \cite{MukhGar,MukhVik}. In this short talk we review our results from the paper \cite{we} with the main stress on  the issues of causality and acoustic metric in eikonal approximation. In addition we correct the formula for the redshift of sound signals from our original work \cite{we}.
In our paper \cite{we} we considered a scalar field $\phi$ with the generally covariant and Lorentz invariant action
\begin{equation}
S_{  }=\int \textrm{d}^{4}x\sqrt{-g}p(X),~~\textrm{where}~~p(X)=\alpha ^{2}\left[ \sqrt{1+\frac{2X}{\alpha ^{2}}}-1\right].
\label{Lagrange}
\end{equation}
The Lagrangian $p(X)$ depends only on $X\equiv \frac{1}{2}\nabla _{\mu }\phi \nabla ^{\mu }\phi$, and $\alpha $ is a free parameter of the theory.
\footnote{Throughout the paper $\nabla _{\mu }$ denotes the covariant derivative and we use the natural units in which $G=\hbar =c=1$.} The kinetic part of the
action is the same as in \cite{MukhVik} and for small derivatives, that is,
in the limit $2X\ll \alpha ^{2},$ it describes the usual massless free
scalar field. In the case of arbitrary $p(X)$ the equation of motion for $\phi$ is

\begin{equation}
G^{\mu \nu }\nabla _{\mu }\nabla _{\nu }\phi =0,
~~\textrm{where the induced metric}~~G^{\mu \nu }\equiv g^{\mu \nu }+\frac{p_{,XX}}{p_{,X}} \nabla ^{\mu }\phi \nabla ^{\nu}\phi,  
\label{general eom}
\end{equation}

and $p_{,X}\equiv \partial p/\partial X$. This equation is hyperbolic and
its solutions are stable with respect to high frequency perturbations
provided $(1+2Xp_{,XX}/p_{,X})>0$ \cite{MukhGar,ArmenLim,Rendall,Komar}.
The propagation vectors $N^\mu$ are tangent to characteristic surface and define the influence cone:
\begin{equation}
G_{\mu \nu }^{-1} N^{\mu} N^{\nu}=0,~~\textrm{where}~~G_{\mu \nu }^{-1}=g_{\mu \nu }-\frac{p_{,XX}}{p_{,X}+2Xp_{,XX}} \nabla_{\mu}\phi \nabla_{\nu}\phi 
\label{cone}
\end{equation}
is inverse matrix to $G^{\mu \nu }$. The influence cone is larger than the light cone if $p_{,XX}/p_{,X}<0$ \cite{ArmenLim,Rendall,Komar}. In this case the front of small perturbations of $\phi$ propagates faster than light. If the background $\phi(x)$ is trivial, $\nabla_{\mu }\phi=0$, then perturbations (small discontinuities) propagate with the speed of light. Therefore only nontrivial backgrounds $\phi(x)$ spontaneously break the Lorentz invariance. Despite the fact that the action (\ref{Lagrange}) is manifestly Lorentz invariant the action for perturbations $\delta \phi$ around a non-trivial background solution is only generally covariant but not Lorentz invariant anymore. This background can be considered as a medium or "new ether".  Observers moving differently with respect to this medium may disagree in the results of some measurements. Moreover in the case of superluminal propagation there is no Lorentz invariant notion of causality \cite{Komar,ArkHamDubov}. However, by virtue of the hyperbolicity of the system even in this case there may exist some Cauchy hypersurfaces \cite{Rendall} and therefore observers for which the causality is well defined \cite{Bruneton}. Nevertheless, there are backgrounds \cite{ArkHamDubov} where closed time like curves (CTC) exist. However, in the standard GR\cite{Bonnor} it is also the case. The so-called chronology protection conjecture \cite{Chronology} may preclude the existence of CTC. For a more detailed discussion of causality in the theories with spontaneously broken Lorentz-invariance see paper \cite{Causality}.
For the energy-momentum tensor we have $T_{\mu\nu}=p_{,X}\nabla_{\mu }\phi \nabla_{\nu}\phi-pg_{\mu\nu}$. Thus the Null Energy Condition $T_{\mu\nu}n^\mu n^\nu\ge0$ \footnote{$n^\mu$ is null vector in $g_{\mu\nu}$} is satisfied if $p_{,X}\ge0$ . This is always the case for our model (\ref{Lagrange}) and hence the black hole
area theorem \cite{Hawking} holds.
It is well known that, if $\nabla _{\nu }\phi $ is timelike (that is, $X>0$ in our
convention), then the system with general $p(X)$ is formally
equivalent to a perfect fluid with the pressure $p=p(X)$, energy density $\varepsilon(X)=2Xp_{,X}(X)-p(X)$, the four-velocity
$u_{\mu }=\nabla _{\mu }\phi/{\sqrt{2X}}$ and the sound speed $c_{s}^{2}\equiv \partial p/\partial \varepsilon=p_{,X}/\varepsilon_{,X}$. Specializing to the case of the Lagrangian (\ref{Lagrange}) we have
\begin{equation}
c_{s}^{2}=1+\frac{2X}{\alpha ^{2}}\ge1,~~\frac{\varepsilon}{\alpha ^{2}} =(1-c_{s}^{-1}),~~\frac{p}{\alpha ^{2}}=(c_{s}-1),~~ G_{\mu \nu }^{-1}=g_{\mu \nu }+\frac{\nabla_{\mu }\phi \nabla_{\nu}\phi}{\alpha ^{2}}. \label{enpres_c}
\end{equation}%
Here we sketch how to find a stationary spherically symmetric background solution
for the scalar field falling onto a Schwarzschild black hole. In the
Eddington-Finkelstein coordinates \footnote{note that these coordinates are regular at Schwarzschild horizon} the metric takes the form:
\begin{equation}
\textrm{d}s^{2}=f(r)\textrm{d}V^{2}-2\textrm{d}V\textrm{d}r-r^{2}\textrm{d}\Omega^{2},~~\textrm{where}~~f(r)\equiv 1-\frac{r_{g}}{r},~~r_{g}\equiv 2M.  \label{metric}
\end{equation}%
In \cite{we} we verified that there is a  broad range of free parameter $\alpha ^{2}$ for which the infalling field has a negligible influence on the black hole, that is, we
consider an accretion of the test fluid in the given gravitational field. 
The stationarity and cosmological boundary conditions at spatial infinity imply the following ansatz for the solution:
\begin{equation}
\phi (V,r)=\alpha \sqrt{c_i^{2}-1}\left( V+\int^r F(r')dr'\right),~\textrm{where}~c_i~\textrm{is the speed of sound at infinity.}
\label{anz}
\end{equation}%
For every solution $F(r)$ the induced acoustic line element (in eikonal approximation) is 
\begin{equation}
\textrm{d}S^{2}\equiv G_{\mu \nu }^{-1}\textrm{d}x^{\mu}\textrm{d}x^{\nu}=\left(c_i^2-\frac{r_{g}}{r} \right)\textrm{d}V^{2}-2\textrm{d}V\textrm{d}r\left(1-(c_i^2-1)F\right)+(c_i^2-1)F^2\textrm{d}r^{2}-r^{2}\textrm{d}\Omega^{2}.  
\label{acoustic metric}
\end{equation} 
In this acoustic metric the coordinate $V$ is timelike. Therefore from (\ref{acoustic metric})  it follows that there exists \textit{sonic horizon} at $r_{\star}=r_g/c_i^2 \le r_g$.
Substituting (\ref{anz}) into (\ref{general eom}) and (\ref{acoustic metric}) one can obtain that the only physical solution which satisfies all boundary conditions and for which the acoustic space-time (\ref{acoustic metric}) is not singular for $r\ge r_{\star}$, is given by:
\begin{equation}
F(x)=\frac{1}{f(x)}\left(\sqrt{\frac{c_{i}^{2}+f(x)-1}{f(x)x^{4}c_{i}^{8}+\left( c_{i}^{2}-1\right) }}-1\right),~~\textrm{where}~~x\equiv r/r_{g}.  \label{F}
\end{equation}
And for the sound speed one obtains $c_{s}^2(x)=x^{3}c_i^{8}/\left[1+c_i^{2}(x-1)(1+xc_i^{2}+x^{2}c_i^{4})\right]$.
The acoustic spacetime (\ref{acoustic metric}) with the function (\ref{F}) describes an analogue black hole with the horizon which is \textit{inside} the Schwarzschild horizon. Therefore it is possible to use perturbations $\delta \phi$ around this background (\ref{anz}),(\ref{F})  as signals and to send information from the region $r_{\star}<r<r_g$ between two horizons, see Fig. 2 from \cite{we}. On the background solution $\nabla ^{\mu }\phi$ is a time like non-vanishing vector field well defined for $r \ge r_{\star}$. Thus in accordance with \cite{Wald} the acoustic space time is stably (and therefore strongly) causal for $r \ge r_{\star}$. 
Suppose that a spacecraft moves together with the falling background field and sends the
acoustic signals with the frequency $\omega _{em}$. After simple
calculations one can obtain that an observer at
rest at the spatial infinity will detect these signals at the frequency $\omega_{inf}$:
\begin{equation}
\frac{\omega_{inf}}{\omega_{em}}=\left(1-\left(\frac{r_{\star}}{r}\right)^{2}\right)\sqrt{\frac{1-r_{g}/r}{1-c_{s}^{2}(r)\left(r_{\star}/r\right)^{4}}}
\end{equation}
This expression corrects our result from \cite{we}. Note that the ratio $\omega _{em }/\omega _{inf}$ is finite for any $%
r>r_{\star }$ and it vanishes for $r=r_{\star}$. In particular for the moment of crossing the Schwarzschild horizon we have $\omega _{em }/\omega _{inf}=c_i^4 \sqrt{1+c_i^2+c_i^4+c_i^6} /(c_i^4-1)$.

\section*{Acknowledgements}

We are very thankful to C.~Bonvin, C.~Caprini, S.~Dubovsky, 
R.~Durrer, V.~Frolov, S.~Liberati, A.~Rendall, S.~Sibiryakov,
 A.~Starobinsky, L.~Susskind, R.~Woodard and especially Sergei Winitzki
 for very useful discussions. A.~V. would like to thank the theory group of Laboratori Nazionali del Gran Sasso, INFN and organizers and staff of Les Houches Summer School for hospitality during the preparation of this manuscript. E.B. thanks Alexander von Humboldt foundation for support on the early stage of this project and INFN for support during the preparation of the manuscript.

\vfill


\end{document}